

\magnification=1200
\voffset=1.5truecm
\font\tit=cmbx12 scaled\magstep 1
\font\abst=cmsl9

\font\aun=cmbx10
\font\rauth=cmcsc10
\font\subs=cmbxti10
\topskip=1truecm
\hsize=6truein
\vsize=8.5truein
\newcount\who
\who=0

\def\centra#1{\vbox{\rightskip=0pt plus1fill\leftskip=0pt plus1fill #1}}

\def\title#1{\baselineskip=20truept\parindent=0pt\centra{\tit #1}
\bigskip\baselineskip=12pt\centra{by}\def\titolo{#1}}

\def\short#1{\def\titolo{#1}}

\def\first#1#2{{\aun #1$^{#2}$}\global\def\firstauthor{#1 {\sl et al.}}}
\def\etal#1#2{, {\aun #1$^{#2}$}}
\def\last#1#2{ and {\aun #1$^{#2}$}}
\def\authors#1{\bigskip\centra{#1}}
\long\def\addresses#1{\bigskip\centra{#1}}
\long\def\addr#1#2{$^{#1}${\it #2}\par}
\long\def\support#1#2{\footnote{}{\hbox to 15truept{\hfill$^{#1}$\ }\sl #2}}
\def\speaker#1{\def\speak{{\sl Talk given by #1}}}
\parskip=5pt
\long\def\summary#1{\bigskip\centra{\speak}\bigskip\medskip
\vbox{\par\leftskip=30truept\rightskip=30truept
\noindent{\bf Summary:} \abst #1}\parindent=15truept}
\def\section#1#2{\who=1\bigskip\medskip\goodbreak{\bf\noindent\hbox to
15truept{#1.\hfil} #2}\nobreak
\medskip\nobreak\who=0}
\def\subsection#1#2{\ifnum\who=0\bigskip\goodbreak\else\smallskip\fi
{\subs\noindent\hbox to 25truept{#1.\hfil} #2}\nobreak
\ifnum\who=0\medskip\fi\nobreak}
\def\references{\bigskip\medskip\goodbreak{\bf\noindent\hskip25truept
References}\nobreak
\medskip\nobreak
\frenchspacing\pretolerance=2000\parindent=25truept}
\def\paper#1#2#3#4#5#6{\item{\hbox to 20truept{[#1]\hfill}}
{\rauth #2,} {\it #3} {\bf #4} (#5) #6\smallskip}
\def\book#1#2#3#4#5{\item{\hbox to 20truept{[#1]\hfill}} {\rauth #2,} {\it #3},
#4 (#5)\smallskip}

\def\Fig#1#2{\medskip\noindent
\hrule\line{\vrule\vbox to#2truemm{\vfill}\hfill\vrule}\hrule\noindent{\bf
Fig.~#1}\medskip}

\newcount\firstp
\firstp=\pageno

\headline={\ifnum\pageno=\firstp\hfill\else
\line{\firstauthor,\quad{\it \titolo}\hfill\folio}\fi}
\footline={\line{\hfill}}


\def\partder#1#2{ {\partial #1 \over \partial #2} }
\def\primato#1{{{#1}^\prime}}
\def\SSs#1{{^{^{#1}}}}
\def\gTRE{\hbox{${^3}\! g$}}

\def\frac#1#2{{ #1 \over #2 }}

\def\ref#1{{(#1)}}

\def\underline#1{{\it #1 }}

\def\EL{\hbox{$\cal E\! L$}}

\def\Ac{{\cal A}}

\def\Fc{{\cal F}}

\def\Hc{{\cal H}}

\def\Lc{{\cal L}}


\title{GALILEAN THEORIES OF GRAVITATION}

\short{Talks given at the X Italian Relativity Conference}

\authors{\first{Roberto De Pietri}{1}\etal{Luca Lusanna}{2}
         \last{Massimo Pauri}{1}}

\addresses{
\addr{1}{ Universit\`a di Parma - INFN, Gruppo Collegato di Parma,
          Parma, Italy}
\addr{2}{I.N.F.N., Sezione di FIRENZE, Arcetri (FI), Italy}
}

\speaker{Roberto De Pietri}
\support{\flat}{Supported by INFN: Iniziativa specifica FI2}

\summary{  A generalization of Newtonian gravitation theory
is obtained by a suitable limiting procedure from the ADM action
of general relativity coupled to a mass-point.
Three particular theories are discussed and it is found that
two of them are invariant under an extended Galilei {\it gauge} group.}

\section{1}{The general framework}

Assuming the existence of a global 3+1 splitting of the space-time
manifold, we consider the Einstein-Hilbert-De Witt action for the
gravitational field plus a matter action corresponding to a
single mass-point:
$$
\eqalign{
\Ac = \Ac_F + \Ac_M
      &=\frac{c^3}{16\pi G} \int dtd^3\! z \sqrt{\gTRE} N
                \left[ { {^3\!}R
                        + \gTRE^{ik} \gTRE^{jl} (K_{ij}K_{kl} - K_{ik}K_{jl})
                        } \right]  \cr
      &~- m c \int d\lambda
                   \sqrt{ - g_{\mu\nu} \primato{x}^\mu \primato{x}^\nu } ~~,
}
\eqno{(1.1)}
$$
where we have adopted the standard ADM notations$^{[1]}$ for the field part.

We want to do the non-relativistic limit of (1.1) in its most general
fashion. By parametrizing the covariant metric tensor as:
$$
 {^4\!}g_{\mu\nu}
           \equiv \left|
              \matrix{ -N^2+ \gTRE^{ij} N_i N_j  &  N_j   \cr
                           N_i              & \gTRE_{ij} \cr }
               \right|
           \equiv - c^2  t_{\mu} t_{\nu} + \tilde{h}_{\mu\nu}
            + O\left( \frac{1}{c^2} \right)
{}~,
\eqno{(1.2)}
$$
$(c=\hbox{velocity of light})$, the inverse metric takes the form:
$$
{^4\!}g^{\mu\nu} \equiv \left|
              \matrix{ - \frac{1}{N^2}   & \frac{\gTRE^{jk} N_k}{N^2}  \cr
                      \frac{\gTRE^{il}N^l}{N^2} &\gTRE^{ij}
                            - \frac{\gTRE^{il}N_l \gTRE^{jk}N_k}{N^2}  \cr }
               \right|
            \equiv h^{\mu\nu} - \frac{1}{c^2} \kappa^{\mu\nu}
             + O\left( \frac{1}{c^4} \right)
{}~~.
\eqno{(1.3)}
$$
{}From the relation
${^4\!}g^{\mu\nu}  {^4\!}g_{\nu\rho} = \delta^\mu_\rho $,
we see that, at first order in $1/c^2$, the following identities
are fulfilled:
$$
 h^{\mu\nu}t_{\nu} = 0 \qquad , \qquad
 h^{\mu\nu}\tilde{h}_{\nu\rho}  = \delta^\mu_\rho
                                  -t_{\rho} \kappa^{\mu\nu} t_{\nu} ~.
\eqno{(1.4)}
$$
Accordingly, in the limit $c\rightarrow +\infty$, we must identify
$h^{\mu\nu}$ with so-called Newtonian {\it space metric} and
$t_{\mu}t_{\nu}$  with so-called Newtonian {\it time metric}
(see for example Refs.[2],[3],[4]).

On the basis of the geometrical assumption made above, we have
to admit the existence of an {\it absolute time} $(T)$ foliation of space-time
and of a global coordinate system in which $t_\mu = T_{,\mu}=(\Theta
(t),\vec{0})$.
In such a system, $g_{\mu\nu}$ takes the form:
$$
\eqalign{
^4\! g_{\mu\nu} &= - c^2 T_{,\mu} T_{\nu} + \tilde{h}_{\mu\nu} \cr
           &\equiv - c^2 \left| \matrix{ \Theta^2 &  0 \cr
                                       0     &  0  } \right|
              +\left| \matrix{ 2 A_0 & A_j    \cr
                                A_i  & g_{ij}  } \right|
              +{1 \over c^2}
                \left| \matrix{ 2 \alpha_0 & \alpha_j    \cr
                                  \alpha_i  & \gamma_{ij}  } \right|
              +{1 \over c^4}
                \left| \matrix{ 2 \beta_0 & \beta_j    \cr
                                  \beta_i & \beta_{ij}  } \right|
             + O\left( \frac{1}{c^6} \right) ~,
}
\eqno{(1.5)}
$$
and
$$
^4\! g^{\mu\nu}
    = \left| \matrix{ 0 &  0      \cr
                      0 &  g^{ij}  }
       \right|
    -{1 \over c^2 \Theta^2}
       \left| \matrix{  1  & - g^{jk} A_k    \cr
             -g^{ik} A_k & g^{ik}g^{jl} ( \gamma_{kl} + A_k A_l)  }
       \right|
    + O\left( \frac{1}{c^4} \right) ~.
\eqno{(1.6)}
$$
Thus, we are lead to identify the ADM variables as:
$$
\left\{.
\eqalign{
  \gTRE_{ij} &\equiv g_{ij} + \frac{1}{c^2} \gamma_{ij}+ \frac{1}{c^4}
\beta_{ij}
                            + O\left( \frac{1}{c^6} \right)   \cr
   {^3\!}R   &\equiv R + \frac{1}{c^2} R_1(g_{ij},\gamma_{ij})
                    +\frac{1}{c^4} R_2(g_{ij},\gamma_{ij},\beta_{ij})
                    + O\left( \frac{1}{c^6} \right)   \cr
   N_i       &\equiv A_i  + \frac{1}{c^2} \alpha_i+ \frac{1}{c^4} \beta_i
                          + O\left( \frac{1}{c^6} \right)   \cr
   N^2       &\equiv c^2 \Theta^2 - 2 A
                + \frac{2}{c^2} \left[{
                          \alpha_0 - g^{ij} \alpha_i A_j
                        - \frac{1}{2} \gamma_{rs} g^{ri} g^{sj} A_i A_j
                          }\right]
                + O\left( \frac{1}{c^4} \right)     \cr
   N K_{ij}  &\equiv {^3\!}B_{ij} = B_{ij} + \frac{1}{c^2} B_{ij}^{\SSs{(1)}}
                                    + O\left( \frac{1}{c^4} \right)  ~,
}
\right.
\eqno{(1.7)}
$$
where we have defined:
$$
\eqalign{
 A      &=  A_0 -\frac{1}{2}g^{ij}A_iA_j   \cr
 B_{ij} &= \frac{1}{2} [\nabla_i A_{j}+\nabla_j A_i -\partder{g_{ij}}{t} ] \cr
 B_{ij}^{\SSs{(1)}}
        &= \frac{1}{2} [\nabla_i \alpha_{j}+\nabla_j \alpha_i
                        -\partder{\gamma_{ij}}{t}
                        - A_k g^{kl} ( \nabla_j \gamma_{il}
                                      +\nabla_i \gamma_{lj}
                                      -\nabla_l \gamma_{ij} ) ] ~.
}
\eqno{(1.8)}
$$

We shall assume the zero$^{{th}}$-order term in the $1/c^2$ expansion
of Eq.(1.1) as our {\it Galilean Action}. Precisely, we have:
$$
\eqalign{
\tilde{\Ac}
       &\equiv \frac{1}{16\pi G} {\int} dtd^3z \sqrt{g}
                \left[{ \Theta R_2 + \frac{1}{2} g^{ij} \gamma_{ij} \Theta  R_1
                        + \frac{1}{2} g^{ij} \beta_{ij} \Theta R
                }\right.    \cr
       &~  ~~  - \frac{1}{2} \Theta
                          \left( g^{ik} g^{jl} - \frac{1}{2} g^{ij} g^{kl}
\right)
                          \gamma_{ij} \gamma_{kl} R
                        - { A \over \Theta } R_1  \cr
       &~  ~~  - { 1\over \Theta }
                          \left({ \alpha_0 - g^{ij} A_i \alpha_j
                                 +\frac{1}{2} A g^{ij} \gamma_{ij}
                                 +\frac{1}{2} \gamma_{ij} g^{il} g^{jm} A_l A_m
                          }\right) R
                        - \frac{A^2}{2\Theta^3} {R}   \cr
       &~  ~~  + \frac{2}{\Theta}
                          g^{ik} g^{jl} ( B_{ij}B_{kl}^{\SSs{(1)}}
                                         -B_{ik}B_{jl}^{\SSs{(1)}})
                        - \frac{2}{\Theta}
                          g^{ik} g^{jr} \gamma_{rs} g^{sl}
                          (B_{ij}B_{kl} - B_{ik}B_{jl})   \cr
       &~  ~~    \left.{\left.{
                        + \frac{A}{\Theta^3}
                          g^{ik} g^{jl} (B_{ij}B_{kl} - B_{ik}B_{jl})
                       } \right] }\right.   \cr
       &~  +m  {\int} d\lambda {m \over \Theta \primato{t}}
                      \left[ \frac{1}{2}g_{ij}
                              (\primato{x}^i + g^{ik} A_k \primato{t} )
                              (\primato{x}^j + g^{jl} A_l \primato{t} )
                             + A \primato{t} \primato{t} \right]  ~~.
}
\eqno{(1.9)}
$$
We see that 27 fields survive the limiting procedure, namely
$\Theta$, $A$, $A_i$, $g_{ij}$, $\alpha_0$, $\alpha_i$,
$\gamma_{ij}$, $\beta_{ij}$,
where we have used $A$ as independent variable instead of
$A_0$.
\section{2}{Particular cases}
\subsection{2.1}{Post-newtonian parametrization}
The maximum of similarity to Newton's theory is
achieved confining to a post-Newtonian-like
parametrization defined by $\Theta = 1$,
$g_{ij} = \delta_{ij}$, $A_i=0$ and $A=-\varphi$, so that:
$$
{^4}\! g_{\mu\nu} = \left| \matrix{- c^2 - 2 \varphi
                       +\frac{2 \alpha_0}{c^2}& \frac{\alpha_i}{c^2} \cr
                        \frac{\alpha_i}{c^2}  &
                        \delta_{ij} + {\gamma_{ij} \over c^2} \cr}
                           \right|   ~.
\eqno{(2.1)}
$$
The action (1.9) becomes:
$$
\tilde{\Ac}   = ~\frac{1}{16\pi G} {\int} dtd^3z
         \left[ {  \varphi R_1 + R_2 - \frac{1}{2} \delta^{ij} \gamma_{ij} R_1
        + m  \left( \frac{1}{2}\delta_{ij} \dot{x}^i \dot{x}^j
                    -\varphi  \right) \delta^3[{\bf z}-{\bf x}(t)]  }\right] ~.
\eqno{(2.2)}
$$
The constraint analysis shows that the {\it secondary} constraints are
just the Euler-Lagrange field equations, which determine the behaviour
of the fields $\varphi$, $\gamma_{ij}$:
$$
\eqalign{
 \chi_{\varphi} ({\bf z},T)
      &\equiv  \frac{1}{16\pi G} R_1
           - m\delta^3\! [{\bf z}-{\bf x}(t)] \simeq 0    \cr
 \chi^{ij}_\gamma ({\bf z},T)
       &\equiv  \frac{1}{16\pi G}
           \big\{  [\delta^{ir}\delta^{js}-\delta^{ij}\delta^{rs}]
                      \partial_r \partial_s  \varphi    \cr
       &~ +{ 1 \over 2}
               [\delta^{ir}\delta^{js}-\frac{1}{2}\delta^{ij}\delta^{rs}]
               \delta^{ab}
                [ \gamma_{ab,rs} + \gamma_{rs,ab}
                 -\gamma_{ar,sb} - \gamma_{as,rb} ]
                                     \big\} \simeq 0  ~,
}
\eqno{(2.3)}
$$
while the mass-point equations of motion are the standard Newton equations
with potential $\varphi$, i.e., $m \ddot{x}^i = - \delta^{ik}
\partial_k\varphi$.
By performing the contraction $\delta_{ij}\chi^{ij}_\gamma$
and using the constraint $\chi_\varphi \simeq 0$, we obtain
the Poisson equation for the potential $\varphi ({\bf z},t)$, i.e.:
$$
 \delta^{ij} \partial_i \partial_j \varphi ({\bf z},t)
      = 4 \pi G m \delta^3\! [{\bf z} - {\bf x}(t)] ~~.
\eqno{(2.4)}
$$
\subsection{2.2}{A ten-fields theory}
If we keep only the fields that explicitly interact with
the mass-point, i.e.,
if we set $\alpha_0 = \alpha_i = \gamma_{ij} = \beta_{ij} =0 $,
the action (1.9) becomes:
$$
\eqalign{
 \tilde{\Ac} &= \frac{1}{16\pi G} \int dtd^3\! z \sqrt{g}
                \left[ { - \frac{A^2}{2\Theta^3} {R}
                        + \frac{A}{\Theta^3}
                          g^{ik} g^{jl} (B_{ij}B_{kl} - B_{ik}B_{jl})
                       } \right]  \cr
          &~  + m  \int dtd^3\! z  {m \over \Theta }
                       \left[ \frac{1}{2}g_{ij}
                              (\dot{x}^i + g^{ik} A_k )
                              (\dot{x}^j + g^{jl} A_l )
                             + A  \right]
                       \delta [{\bf z} - {\bf x}(t)] ~~. \cr
}
\eqno{(2.5)}
$$
It can be easily shown that the field $\Theta( t)$ lacks real
dynamical content. Actually, its role amounts only to a redefinition of the
evolution parameter $t$ in the expression
$T(t)=\int_0^t d\tau \Theta (\tau )$, which, in turn, has to be identified to
Newtonian {\it absolute time}. Indeed, if we redefine the
fields $A$, $A_0$, $A_i$ and $B_{ij}$ as:
$$
\left\{ {
\eqalign{
\tilde{A}_0    &\equiv {A_0 \over \Theta^2}
              ~~;\qquad  \tilde{A} \equiv {A   \over \Theta^2}
              ~~;\qquad \tilde{A}_i  \equiv {A_i \over \Theta}   \cr
\tilde{B}_{ij} &\equiv {{B}_{ij}  \over \Theta}
                       =  \frac{1}{2} [ \nabla_i \tilde{A}_{j}
                             + \nabla_j \tilde{A}_i
                             - \partder{g_{ij}}{T} ] ~~,
}
}\right.
\eqno{(2.6)}
$$
the total action (2.5) can be rewritten in the form:
$$
\eqalign{
  \tilde{\Ac}  &= \frac{1}{16\pi G} \int dTd^3\! z \sqrt{g}
                \left[ { - \frac{\tilde{A}^2}{2} {R}
                        + \tilde{A}
                          g^{ik} g^{jl} (\tilde{B}_{ij}\tilde{B}_{kl}
                                        -\tilde{B}_{ik}\tilde{B}_{jl})
                       } \right]  \cr
          &~~+ m  \int dTd^3\! z
                       \left[ \frac{1}{2}g_{ij}
                              ( \frac{d{x}^i}{dT} + g^{ik} \tilde{A}_k )
                              ( \frac{d{x}^j}{dT} + g^{jl} \tilde{A}_l )
                             + \tilde{A}  \right]
                      \delta [{\bf z} - {\bf x}(T)] ~~, \cr
}
\eqno{(2.7)}
$$
where $\Theta$ has disappeared. Then, the Dirac Hamiltonian is:
$$
\eqalign{
 H_D &= \int d^3\!z \left\{ \frac{ \tilde{A}^2 }{ 32\pi G} \Hc_I
                           +\frac{ 16\pi G}{ \tilde{A} } \Hc_E
                    +\left[ \frac{1}{2m} g^{ij} p_i p_j - m \tilde{A} \right]
                     \delta [{\bf z} - {\bf x}(T)] \right\} \cr
      &~+ \int d^3\!z  \left[ \tilde{A}_i g^{ij} \phi_j
                             +\pi^i \lambda_i + \pi_A \lambda_A \right]  ~,
}
\eqno{(2.8)}
$$
where $\pi^{ij}$, $\pi^i$, $\pi_A$ are the conjugate fields of
$g_{ij}$, $\tilde{A}_i$, $\tilde{A}$, respectively, and
$\Hc_I = \sqrt{g} R$,
$\Hc_E = \frac{1}{\sqrt{g}} [{\rm Tr}\pi^2-\frac{1}{2} ({\rm Tr} \pi)^2]$,
$\phi_j = -2 g_{jk} \nabla_l \pi^{kl} - p_j \delta [{\bf z} - {\bf x}(T)]$.
The constraints' chains are:
\Fig{ten-fields constraints' chains}{54}
\noindent
where the arrows denote requirement of time-conservation of the
constraints\footnote{$^\star$}{The explicit expression of the constraints
can be found in Ref.[5]}\rlap .
Thus, the fields $A_i$ and {\it three} independent functionals of $g_{ij}$
are {\it gauge} variables, while $\tilde{A}$ is determined by
the constraint $\chi_A\simeq 0$. The surviving physical degrees of freedom
correspond to {\it three} independent functionals of $g_{ij}$.

It is worth considering the particular solution of the variational problem
corresponding to the {\it static sector}, defined by
$A_k = 0$, $\frac{dg_{ij}}{dT}=0$, $\frac{dx^i}{dT}=0$. The consistency of the
constraints' chains, in this case, implies
that the field $\tilde{A}$ satisfies the following {\it modified
Poisson equation}:
$$
{\sqrt{g} } g^{ij} \nabla_i \nabla_j \tilde{A}^2
       = 4\pi G m \tilde{A} \delta [{\bf z} - {\bf x}(T)] ~.
\eqno{(2.9)}
$$
It is remarkable that, under the assumption that $\tilde{A}=\tilde{A}(r)$ and
$g_{ij} \rightarrow \delta_{ij}$ for
$r=|\vec{z}-\vec{x}(T)|\rightarrow \infty$, the field $\tilde{A}$, at fixed
$T$, has the asymptotic expansion
$$
 \tilde{A} = {k_1 \over r} +  {k_2 \over r^2} + .. ~~.
\eqno{(2.10)}
$$
\subsection{2.3}{An eleven-fields theory}
The action for this case is given by the expression (2.5)
where now $\Theta$ is allowed to be a function depending
also on the space coordinate $\vec{z}$: $\Theta =\Theta (\vec{z},t)$.
This has the effect that the resulting theory has the same number of degrees of
freedom of Einstein's theory. Although  the function  $\Theta (\vec{z},t)$
cannot be rescaled in this case,
the corresponding additional degree of freedom, which is a classical analogue
of the {\it dilaton}, does not have any propagation properties.

The Dirac Hamiltonian is:
$$
\eqalign{
 H_D &= \int d^3\!z \Theta \left\{ \frac{ \tilde{A}^2 }{ 32\pi G} \Hc_I
                           +\frac{ 16\pi G}{ \tilde{A} } \Hc_E
                    +\left[ \frac{1}{2m} g^{ij} p_i p_j - m \tilde{A} \right]
                     \delta [{\bf z} - {\bf x}(T)] \right\} \cr
      &~+ \int d^3\!z  \left[ \tilde{A}_i g^{ij} \phi_j
                             +\pi^i \lambda_i + \pi_A \lambda_A
                             +\pi_\Theta \lambda_\Theta\right] ~,
}
\eqno{(2.11)}
$$
where $\pi^{ij}$, $\pi^i$, $\pi_A$, $\pi_\theta$ are the conjugate fields of
$g_{ij}$, $\tilde{A}_i$, $\tilde{A}$, $\Theta$ , respectively,
and $\Hc_I$, $\Hc_E$, $\phi_j$ have the same functional form as before.
The constraints' chains are now:
\Fig{eleven-fields constraints' chains}{70}
\noindent
The fields $A_i$ and {\it three} independent functionals of $g_{ij}$
are still {\it gauge} variables. On the other hand,
despite the additional degree of
freedom introduced by $\Theta$, we have now two more pairs of
{\it second-class} constraints. Therefore only two dynamical
{\it graviton-like} degrees of freedom survive in the three-metric $g_{ij}$.
\section{3}{Gauging the Extended Galilei group}
Within the re-parametrization
invariant formulation of the non-relativistic free mass-point system
with coordinates $t(\lambda) ,x^i(\lambda) $, the action and the
Lagrangian are:
$$
\tilde{\Ac}_M = \int_{{\lambda}_1}^{{\lambda}_2}
                     d{\lambda}~~{L}_M(\lambda)
            = \int_{{\lambda}_1}^{{\lambda}_2}
                    d{\lambda}~ \frac{1}{2} m
                     {\delta_{ij} \primato{x}^i(\lambda) \primato{x}^j(\lambda)
                      \over \primato{t}(\lambda)}
                        ~~,
\eqno{(3.1)}
$$
where $\primato{f}(\lambda ) \equiv \frac{d}{d{\lambda}} f(\lambda )$.
The infinitesimal Galilei transformations, defined as
``equal-$\lambda$'' ones (i.e., ${\delta} \lambda = 0$),
are:
$$
\left\{
\eqalign{
  {\delta} t   &= - \varepsilon \cr
  {\delta} x^i &= \varepsilon^i+c^{~~i}_{jk}\omega^j x^k -t v^i
                      \equiv \eta^i
\;\; .
}
\right.
\eqno{(3.2)}
$$
Under the transformations (3.2), it follows:
$$
{\delta}{L}_M  = \frac{d}{d{\lambda}} \left[{
                                  {- m \delta_{ij} v^i x^j } }\right]
 ~~,
\eqno{(3.3)}
$$
which shows that the {\it quasi-invariance} of the Lagrangian is an
effect of the {\it central-charge} term.

 The infinitesimal {\it gauge} Galilei transformations are defined by
allowing the following space-time dependence for the infinitesimal parameters
of (3.2): $\varepsilon (t)$, $\varepsilon^i({\bf x},t)$, $\omega^j({\bf x},t)$,
$v^i({\bf x},t)$. Accordingly, the mass-point coordinates
transform as:
$$
\left\{
\eqalign{
\delta t             &= -\varepsilon (t)  \cr
\delta x^i           &= \eta^i ({\bf x},t)
}
\right.
\qquad
\left\{
\eqalign{
\delta \primato{t}   &= - \primato{t}
                         \frac{d\varepsilon (t)}{dt} \cr
\delta \primato{x}^i &=  \primato{x}^k \partder{\eta^i ({\bf x},t)}{x^k}
                         +\primato{t}  \partder{\eta^i ({\bf x},t)}{t} ~.
}
\right.
\eqno{(3.4)}
$$
To save {\it quasi-invariance} of a new putative
matter Lagrangian ${L}_M^g$ under the {\it gauge} transformation
(3.4), {\it compensating} fields must be introduced.
If, in addition, a global flat limit of fields' expressions
and transformation properties (including the {\it central charge term})
is required, the whole set of conditions are satisfied by
the following choices:
$$
{L}_M^{g} (\lambda)
                    \equiv {1\over\Theta\primato{t}} \frac{m}{2}
                           \left[ {  g_{ij} \primato{x}^i  \primato{x}^j
                                    + 2 A_i \primato{x}^i  \primato{t}
                                    + 2 A_0 \primato{t}    \primato{t} }
                           \right] ~,
\eqno{(3.5)}
$$
and
$$
\left\{
\eqalign{
{\delta} \Theta  &= \dot{\epsilon}(t) \Theta (t)  \cr
{\delta} g_{ij}  &= - \partder{\eta^k ({\bf x},t) }{x^i} g_{kj}
                      - \partder{\eta^k ({\bf x},t) }{x^i} g_{kj} \cr
{\delta} A_i     &= \dot{\varepsilon} A_i
                    - A_j \partder{\eta^j}{x^i}
                    - g_{ij}  \partder{\eta^j}{t}
                    + \Theta \partder{\Fc}{x^i} \cr
{\delta} A_0     &=  2 \dot{\varepsilon} A_0
                     - A_i \partder{\eta^i}{t}
                     + \Theta \partder{\Fc}{t} ~~,
}\right.
\eqno{(3.6)}
$$
where $\Fc = - g_{ij} v^i x^j$. Indeed
$$
\delta {L}_M^{g} = m {d \Fc \over d\lambda} ~~.
\eqno{(3.7)}
$$
\section{4}{Local Galilei invariance}
Unlike the theory corresponding to the post-Newtonian
parametrization, both the ten-fields and eleven-fields theories
admit the just defined {\it gauge} Galilei transformations
as a {\it local symmetry}.
Indeed, for both theories, the variation of the total action
under the transformations of the mass-point coordinates and {\it gauge}
fields (3.4),(3.6), is given by~~
($\tilde{\Ac}=\int dt d^3\!z \tilde{\Lc}$ and
$\Gamma ={\rm Tr} B^2 - ({\rm Tr} B)^2$)~:
$$
\eqalign{
{\delta} \tilde{\Ac} &= \int dtd^3\! z
                         \left\{  \dot\varepsilon \tilde{\Lc}
               +\left[ {
                      \frac{1}{16\pi G} \frac{\sqrt{g}}{\Theta^2}
                      \left(  - {A} {R}  + \Gamma \right)
                     + m \delta [{\bf z} - {\bf x}(t)]
                } \right]
     \left[ \partder{\Fc}{t} - A_r g^{rs} \partder{\Fc}{z^s} \right]
                 \right. \cr
           &~~~ -  \frac{1}{8\pi G}
          \Bigg[ \partder{}{z^j} \left( {
                           \sqrt{g} \frac{A}{\Theta^2}
                         [ B^{ij}  - ({\rm Tr} B) g^{ij} ]
                              }\right)
                 + \sqrt{g} \frac{A}{\Theta^2}
                         [ B^{rs}  - ({\rm Tr} B)g^{rs} ] \Gamma^i_{rs}
                 \cr
           &~~~  \left. - { m  \delta [{\bf z} - {\bf x}(t)]
                      ~\left( \dot{x}^i + g^{ij} A_j \right)
                      } \Bigg] \partder{\Fc}{z^i}
               + \frac{1}{8\pi G}  \partder{}{z^i}
                        \left( \frac{\sqrt{g} A}{\Theta^2}
                               [B^{ij} - ({\rm Tr}B) g^{ij} ]
                               \partder{\Fc}{z^j}
                             \right) \right\} \cr
           &= \int dt d^3\! z \left\{ { \dot\varepsilon \tilde{\Lc}
                     +  \Theta \EL_{A}
               \left( \partder{\Fc}{t} - A_r g^{rs} \partder{\Fc}{z^s}
                    \right)
              +\Theta \EL_{A_i} \partder{\Fc}{z^i}   }\right. \cr
           &~~~ \left. {
              +\frac{1}{8\pi G} \partder{}{z^i}
                        \left( \frac{\sqrt{g} A}{\Theta^2}
                               [B^{ij} - ({\rm Tr}B) g^{ij} ]
                               \partder{\Fc}{z^j}
                               \right)  } \right\} ~~,
}
\eqno{(4.1)}
$$
where $\EL_{A}$ and $\EL_{A_i}$ are the Euler-Lagrange derivatives
that are zero on the extremals.
In conclusion, the total action is {\it quasi-invariant} under
the considered transformations {\it in force} of the equations of motion.
This feature is precisely what it should
be expected in the case of a variational principle corresponding to
a singular Lagrangian.


\references

\book{1}{Arnowitz, Deser, C.W. Misner}{The dynamics of general
relativity}{in: Gravitation: an introduction to current
research}{John Wiley \& Sons, New York 1962}

\paper{2}{K. Kucha\v{r}}{Phys. Rev. D}{22}{1980}{1285}

\paper{3}{H.P. K\"{u}nzle}{Ann. Inst. Henry Poincar\'e}{42}{1972}{337}

\paper{4}{G. Dautcourt}{Acta Phys. Pol.}{25}{1964}{637}

\book{5}{R. De Pietri, L. Lusanna and M. Pauri}{Generalized Newtonian
gravities as ``gauge'' theories of the extended
Galilei Group}{University of Parma preprint UPRF-92-346}{1992}

\bye